\documentclass[nologo,11pt,a4paper]{ETHpaper}

\usepackage{tabularx}
\usepackage{MnSymbol}
\usepackage{booktabs}
\usepackage{caption}
\usepackage{hhline}
\usepackage[table,dvipsnames]{xcolor}
\usepackage{graphicx}
\usepackage[english]{babel}
\usepackage[numbers,sort&compress]{natbib} 
\usepackage{setspace}
\usepackage{microtype}

\title{Social percolation revisited: \\ From 2d lattices to adaptive networks}
\titlealternative{Social percolation revisited: From 2d lattices to adaptive networks} 
\author{Frank Schweitzer}
\authoralternative{Frank Schweitzer}
\address{Chair of Systems Design, ETH Zurich, Switzerland\\
  \url{www.sg.ethz.ch}}
\www{\url{http://www.sg.ethz.ch}}

\reference{\emph{Submitted for publication in Physica A}} 
\makeframing

\usepackage{appendix}
\usepackage{amsmath} \usepackage[utf8]{inputenc}
\usepackage{xcolor} \usepackage{soul}
\usepackage[hang]{subfigure}
\usepackage{multirow}
\usepackage{hyperref}
\usepackage{xparse,tikz,calc} \usepackage{pifont}

\renewcommand{\epsilon}{\varepsilon}

\newcommand{\mean}[1]{\left\langle #1 \right\rangle}
\newcommand{\abs}[1]{\left| #1 \right|}

\begin{document}

\makeframing
\maketitle

\begin{center}
  \emph{\large In memory of Dietrich Stauffer}
\end{center}

\begin{abstract}
  The social percolation model \citep{solomon-et-00} considers a 2-dimensional regular lattice.
  Each site is occupied by an agent with a preference $x_{i}$ sampled from a uniform distribution $U[0,1]$.
  Agents transfer the information about the quality $q$ of a movie to their neighbors only if $x_{i}\leq q$.
  Information percolates through the lattice if $q=q_{c}=0.593$. --
  From a network perspective the percolating cluster can be seen as a random-regular network with $n_{c}$ nodes and a mean degree that depends on $q_{c}$.
  Preserving these quantities of the random-regular network, a true random network can be generated from the $G(n,p)$ model after determining the link probability $p$.
  I then demonstrate how this random network can be transformed into a threshold network, where agents create links dependent on their $x_{i}$ values. 
  Assuming a dynamics of the $x_{i}$ and a mechanism of group formation, I further extend the model toward an adaptive social network model. 
  \end{abstract}

\section{Introduction}
\label{sec:introduction}

Dietrich Stauffer (DS) had a persistent interest in models of social dynamics.
Therefore, he became an early member and a strong supporter of the former working group (now division) \emph{``Physics of Socio-Economic Systems''} ($\Phi$SOE) of the German Physical Society, which I founded together with Dirk Helbing in 2001.\footnote{\url{https://www.dpg-physik.de/vereinigungen/fachlich/skm/fvsoe/aims-and-scope}}
Two books by DS and co-authors, \emph{``Evolution, Money, War and Computers''} (1999) \citep{de_Oliveira_1999} and 
\emph{``Biology, Sociology, Geology by Computational Physicists''} (2006) \citep{Stauffer_2006} nicely demonstrate his approach towards socio-economic topics.
I think that this approach is still well characterized by what we wrote in a review at that time:
{\sl ``Common to both books, statistical physics and computer simulations are creatively used to explain the most diverse phenomena.
(...) 
Albeit, in-depth analysis of the topics discussed is not the authors intention. They rather prefer a reductionistic, sometimes a somewhat superficial way of presenting the problem under consideration. (...)
In many cases, this seems to be a good starting point for further investigations, and scientists obsessed with the idea of putting as much as possible into their computer simulations could be impressed by the results obtained already from basic models. (...)
''}
\footnote{\url{http://jasss.soc.surrey.ac.uk/10/1/reviews/schweitzer.html}}

For this short paper, I focus on one particular publication of DS with relevance for my own research field: \emph{Social percolation models} \citep{solomon-et-00}, henceforth referenced as the SP paper, or SP model. 
I remember that DS explained this model to me during a summer school in Rovinj (Croatia), we jointly organized for the German Academic Scholarship Foundation in 2002.
The SP paper very much follows the spirit described for the books.
Still, it received considerable attention: 
272 Google citations and 138 citations according to the Web of Science (September 2020).

The main idea of the SP paper is to study the ``hit and flop'' dynamics observed for products such as movies  (there is also a follow-up: \citep{Weisbuch_2000}).
The authors' explanation, in a nutshell: a movie becomes a ``hit'' if the information about it can percolate through a lattice, and a ``flop'' otherwise.
To reach the percolation threshold, the movie needs to have a quality $q=q_{c}$, where $q_{c}=0.59$ on a 2d lattice.
In the following, I will summarize the main ideas and comment on possible extensions, to make the social percolation model a ``social'' one.

\section{From lattice structures to random networks}
\label{sec:from-latt-struct}

\subsection{Social percolation on a lattice}
\label{sec:model}

Let us consider a multi-agent system, where each agent $i=1,...,N$  can directly influence a number of other agents, $k_{i}$. 
Taking the complex network perspective, agents are represented by nodes and their interactions by links.
So, we have in total $N$ nodes and $m$ links. $k_{i}$ is known as the \emph{degree} of an agent.

Instead of a network the SP paper discusses the \emph{regular topology} of lattices.
In a 2d regular lattice of size $L^{2}$ each lattice site is connected to 4 neighboring sites by a \emph{bond}.
If each site is occupied by an agent, we have $N=L^{2}$ agents in total.
To introduce the process that lend its name to the SP paper, the authors assume that each agent is characterized by a scalar variable $x_{i}\in [0,1]$ which is sampled from a \emph{uniform distribution}, $U(x)=U[0,1]$.
$x_{i}$ denotes an agent's individual ``taste'', ``quality expectation'', or preference in general.
In the first version of their model, $x_{i}$ is constant over time.
Initially, a few agents get exposed to a movie that has a quality $q \in[0,1]$.
If the movie quality exceeds their expectation, $x_{i}\leq q$, these agents spread this information to their 4 lattice neighbors.
If instead $x_{i}>q$, they do \emph{not} spread any information.
Agents that have received information about $q$ from their neighbors will compare it with their own preference, $x_{k}$.
Only if $x_{k}\leq q$, they spread the information further, and so forth.
As the result of this process, the SP paper finds that agents with $x_{i}\leq q$ will form a spanning cluster only if the movie quality is above a critical threshold, $q\geq q_{c}$.
This is just another interpretation of the phenomenon of \emph{percolation}, and DS made significant contributions to percolation theory \citep{Stauffer_1979}. 

That means, in the SP model we distinguish two types of agents, those that actively contribute to the spread of information and those that don't.
Only the former are of interest, as they form the percolating cluster. 
Let us introduce the density $\rho(q)=n(x_{i}\leq q)/L^{2}$, where $n(x_{i}\leq q)$ is the total number of agents with a preference $x_{i}$ smaller or equal to a given value $q$.
Because $U(x)$ is a uniform distribution, the density is given as $\rho(q)=\int_{0}^{q}U(\hat{x})d\hat{x}=q$.
I.e. $q$ defines the fraction of agents that will successfully spread the information about the movie to their neighbors.

\begin{figure}[htbp]
  \includegraphics[width=0.35\textwidth]{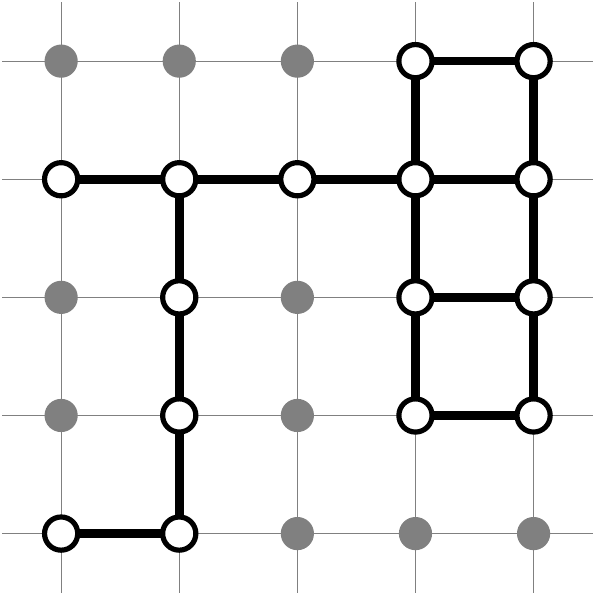}(a)
  \hfill
  \raisebox{1ex}{\includegraphics[width=0.36\textwidth]{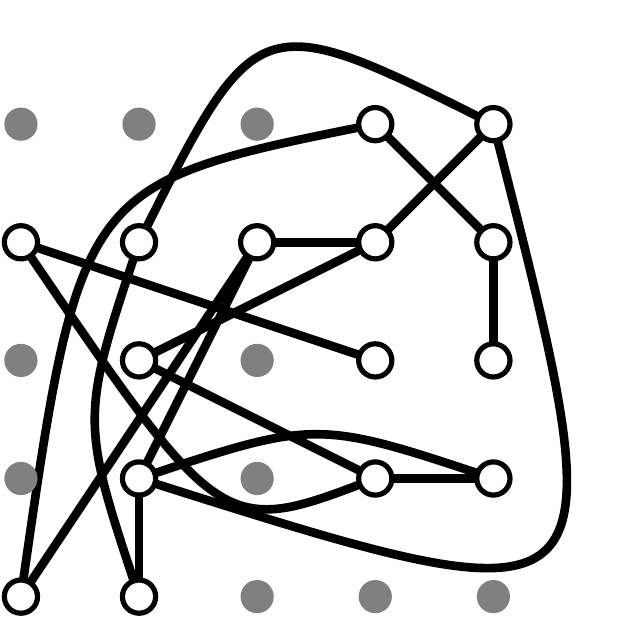}}(b)
\caption{(a) Sketch of a 2d lattice and the network resulting from $\rho^{s}$=0.593. (b) Sketch of a random network with the same density. For gray nodes $x_{i}>q$, for white nodes $x_{i}\leq q$.
    }
  \label{fig:2d}
\end{figure}

It is known for 2d regular lattices that percolation occurs if $\rho\equiv\rho_{c}=0.593$, which is the threshold for  \emph{site} percolation.
For a given lattice size $L^{2}$, the value $\rho_{c}=q_{c}$ defines the minimum number of agents $n_{c}$ with a preference  $x_{i}\leq q_{c}$ that form the percolating cluster. 
This cluster can be seen as a \emph{network} on the underlying lattice, as illustrated in Figure~\ref{fig:2d}(a). 
The network size is $n_{c}$ instead of $N$ because agents with $x_{i}> q$ by construction cannot be part of the network.
For the sample network with $L=5$ shown in Figure~\ref{fig:2d}, we obtain $n_{c}=15$ for the number of nodes. 
For the value $L=4000$ used in the SP paper, we get roughly $n_{c}=9.5\times 10^{6}$.

This network has of course no longer a regular topology, but it is not a random network either as constructed by the procedures mentioned below.
So, let us call this a \emph{random-regular network} in the following.

\subsection{Hits and flops}
\label{sec:hits-flops}

We now formalize the outcome of the model, i.e. the measures for hits and flops.
Each agent $i$ is characterized, in addition to $x_{i}$, by a variable $s_{i}(t)$ that describes whether the agent has \emph{adopted} the movie, $s_{i}(t)$=1, or not $s_{i}(t)$=0.
Adoption implies that $x_{i}\leq q$ and that agent $i$ spreads this information to its neighbors in the random network.
So, we introduce \citep{Lorenz2009b}: 
\begin{align}
  \label{eq:1}
  s_{i}(t)=\Theta\left[w_{i}(t)\right]\;; \quad   w_{i}(t)=\phi_{i}(t)-x_{i}
\end{align}
where the Heaviside function is $\Theta[y]=1$ if $y\geq 0$ and zero otherwise.
$\phi_{i}(t)=0$ if the agent has not received any information about the quality of the movie, and $\phi_{i}(t)=q$ if it does.
More precisely,
$\phi_i(t) = q\,\Theta[\sum_{i_{j}} s_{i_{j}}(t-1)-1]$ where $i_{j}$ refers to the four lattice site neighbors $j$ of agent $i$. 

The \emph{emergence} of a percolating cluster can then be measured by monitoring the fraction of agents forming clusters.
There are two ways to define this fraction.
If we choose the lattice size $N$ for reference as in the SP paper, then $X(t)=\frac{1}{N}\sum_{i=1}^{N} s_{i}(t)$.
``Social percolation'' means that for large $t$ the fraction $X$ approaches $q\geq q_{c}$. 
If $q<q_{c}$, the spread of information will only generate small clusters that do not encompass the whole network.
In the first case, the movie is a ``hit'', in the second case it is a ``flop''.
We note that this distinction holds only for large $N$. 
Alternatively, we can also choose as a reference the number $n$ of those agents that fulfill the condition $x_{i}\leq q$, i.e. $\hat{X}(t)=\frac{1}{n}\sum_{i=1}^{n} s_{i}(t)$ \citep{Lorenz2009b}.
Then, social percolation means that over time $\hat{X}\to 1$.

In both cases the outcome is entirely determined by the value of $q$ which is either larger or smaller than $q_{c}=\rho_{c}$. 
From the modeling perspective this should be seen as a drawback. 
Thus, ample ways have been proposed to enrich those models such that nontrivial results are obtained.
Because of the limited space, I can only mention a few of the recent developments.
While the SP model only assumes that a fixed quantity $q$ is transferred between agents, more refined models of load distribution give dynamic expressions for $\phi_{i}(t)$  \citep{Lorenz2009b,Burkholz2018,OLRDspatial,PhysRevE.90.062816}.
While the SP model considers regular lattices, recent models also consider various network topologies \citep{Gleeson2008}.
While the SP paper only presents computer simulations, heterogenous mean-field approximations allow for analytical solutions \citep{Burkholz2018a,Gleeson2007}, even for finite networks \citep{Burkholz2018-2}.
Further, such contagion-type models have been developed into models of systemic risk that allow to quantify the impact of external shocks by calculating $\hat{X}(t)$ \citep{Tessone2013a,Lorenz2009b}.

\subsection{From random-regular to random networks}
\label{sec:from-random-regular}

In a next step, we want to turn the  topology of the random-regular network into a topology of a true random network,  to get rid of the constraints from the underlying 2d lattice.
To make both networks comparable, we have to decide what parameters should be kept constant. 
If we keep the number of nodes in the percolating cluster, $n_{c}$, we still need a second parameter, because networks are defined by nodes and links.
Our choice is to keep the average degree $\mean{k}$.

What is the average degree for the random-regular network?
The probability that a randomly chosen lattice site is occupied by an agent with $x_{i}\leq q$ is given by $\rho$, and only those agents can be part of the network.
Because each agent has four neighbors on the lattice, we have for the random-regular network $\mean{k}=4\rho^{2}$.
For a true random network, on the other hand, we know that $\mean{k}=np$, where $n$ is the size of the network and $p$ is the probability to form a link between any two agents from $n$.
This gives us $p=\mean{k}/n=4\rho^{2}/n$ for the random network.

These values allow to construct a true random network by means of the $G(n,p)$ model.
It creates a link between any pair of agents with a certain probability $p$.
With $\rho=\rho_{c}=0.593$ and $n=n_{c}=\rho_{c}L^{2}$ we find $p_{c}=4\rho_{c}/L^{2}$.
For the sample lattice with $L=5$ this gives $p_{c}=0.094$, and for the lattice of the SP paper $p_{c}=0.148\times 10^{-6}$.
Figure~\ref{fig:2d}(b) shows a true random network obtained from the $G(n,p)$ model, the topology of which can be directly compared to the random-regular network.

\section{From random networks to threshold networks}
\label{sec:from-random-networks}

\subsection{Generating a social network}
\label{sec:gener-soci-netw}

From a social science perspective, a major drawback of the SP model is the preassigned social structure.
The underlying 2d regular lattice is often interpreted in a social manner by assuming that the lattice encodes a spatial neighborhood.
Then links appear \emph{only} between local neighbors.
This leads to a ``social'' network, in which two to four agents from the same neighborhood are linked if their preferences are $x_{i}\leq q$.

How realistic are these assumptions for the social situation assumed, namely to talk about the quality of the movie just watched?
Would you talk to randomly chosen neighbors with a completely different taste, just because they happen to live in your local neighborhood?
You would probably share your experience with your friends on a social network, where \emph{local} neighborhood is replaced by \emph{social} neighborhood. 
But broadcasting the information is precisely \emph{not} considered in the SP model, there are only bilateral interactions possible.

To make the social percolation model a \emph{social} one, we should start by asking \emph{why} do agents create links to other agents.
In a social system, links are not ``there'', they are actively established and can also disappear again, this way leading even to the collapse of large online social networks \citep{Garcia2013b}.

Given the fixed values $n$, $p$ obtained from the random-regular network, as a first social ingredient we assume that agents want to share their experience likely with those who also share their taste. 
Consider that pairs of agents are successively sampled uniformly at random, without replacement.
Agents form a link $a_{ij}$ only if the differences in their preferences are below a certain threshold $\epsilon$.
We define:
\begin{align}
   \label{eq:4}
   a_{ij}=\Theta\left[z_{ij}\right]\;; \quad z_{ij}= \epsilon -\abs{x_{i}-x_{j}}
\end{align}
The $a_{ij}$ have values 0,1 and denote the entries of an \emph{adjacency matrix} $\mathcal{A}$ that completely describes the topology of the social network.
To restrict the interaction by means of $\epsilon$ was originally proposed in the so-called ``bounded confidence'' model, where $x_{i}$ represented the opinion of an agent \citep{hegselmann2002opinion,deffuant2001mixing,Schweitzer2020}.

Obviously, the density of the social network created this way depends on  $\epsilon$.
On the other hand, the density of the true random network is determined by the probability $p$. 
To derive a relation between $\epsilon$ and $p$, we make use of the fact that the $x_{i}$ are uniformly distributed in the interval $[0,1]$.
The percolating cluster only contains agents with $x_{i}\leq q_{c}$.
To map this back to the unit interval, we normalize $\tilde{x}_{i}=x_{i}/q_{c}$. 
For the formation of the social network $\Delta x_{ij}=\abs{x_{j}-x_{i}}$ matters, so
$\Delta \tilde{x}_{ij}=\Delta x_{ij}/q_{c}$. 
The distribution function $P(\Delta \tilde{x})$ of the absolute difference between two uniform variables $\tilde{x}_{i}$, $\tilde{x}_{j}$ is known as the \emph{triangular distribution}.
The cumulative probability $F(\Delta \tilde{x} \leq \epsilon)$ to find a value $\Delta \tilde{x} \leq \epsilon$ follows likewise:
\begin{align}
  \label{eq:5}
  P(\Delta \tilde{x}) =&\ 2 -2 \left(\Delta \tilde{x} \right) \;; \quad 0\leq \Delta \tilde{x} < 1 \nonumber \\
F(\Delta \tilde{x} \leq \epsilon) = &\ 2 \epsilon - \epsilon^{2}
\end{align}
To get back to the preference values $x_{i}$, we use $q_{c}=0.59$ and transform
\begin{align}
  F(\Delta x \leq \epsilon)=F(\Delta \tilde{x} \times q_{c} \leq \epsilon)=F(\epsilon)=2 \left(\frac{\epsilon}{q_{c}}\right)- \left(\frac{\epsilon}{q_{c}} \right)^{2}
\end{align}
$F(\epsilon)$ has a clear interpretation as the fraction of possible pairs of agents in the network that have a link in common. 
The maximum of $F(\epsilon)$ is obtained when $\epsilon=q_{c}$, which means that with $\epsilon\geq q_{c}$ all agents with $x_{i}\leq q_{c}$ are indeed connected. 
In the $G(n,p)$ model, we sample $n(n-1)/2$ pairs of agents and the probability to successfully establish a link is $p$.
Hence, the fraction of pairs with a link is $p$.
To connect the two, we get
\begin{align}
  \label{eq:6}
  p=\frac{\mean{k}}{n}\equiv F(\epsilon)= 2 \left(\frac{\epsilon}{q_{c}}\right)- \left(\frac{\epsilon}{q_{c}} \right)^{2}
\end{align}
This quadratic equation has two roots:
\begin{align}
  \label{eq:7}
  \epsilon= q_{c} \left[ 1 \pm \sqrt{1- p}\right]
\end{align}
Only the lower value makes sense in our model, because  $\epsilon=q_{c}$ already means that all $n_{c}$ agents are connected.
If $p=0$, i.e. if we have  an empty network, $\epsilon=0$.
If $p=1$, i.e. if we have  an fully connected network, $\epsilon=q_{c}$.
For the sample network with $p=0.094$ we find $\epsilon=0.028$.
For the lattice used in the SP paper, we find with $p=0.148\times 10^{-6}$ for $\epsilon=0.438 \times 10^{-7}$. 
That is not a surprise, because with about $10^{7}$ nodes the network is quite large, but still has to be \emph{sparse}.

The importance of Eqn.~\eqref{eq:7} should not be underestimated, because it allows to relate two very different classes of network models, the random network models described by a probability $p$ and the threshold network models described by a tolerance $\epsilon$.
Because $p$ is related to the network density, Eqn.~\eqref{eq:7} can be also used to link generating mechanisms other than the $G(n,p)$ model to threshold models. 

\begin{figure}[htbp]
  \includegraphics[width=0.33\textwidth]{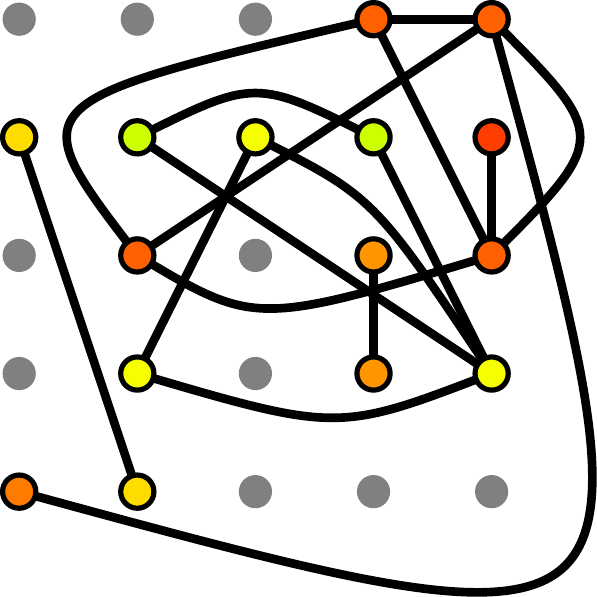}(a)
  \hfill
  \includegraphics[width=0.33\textwidth]{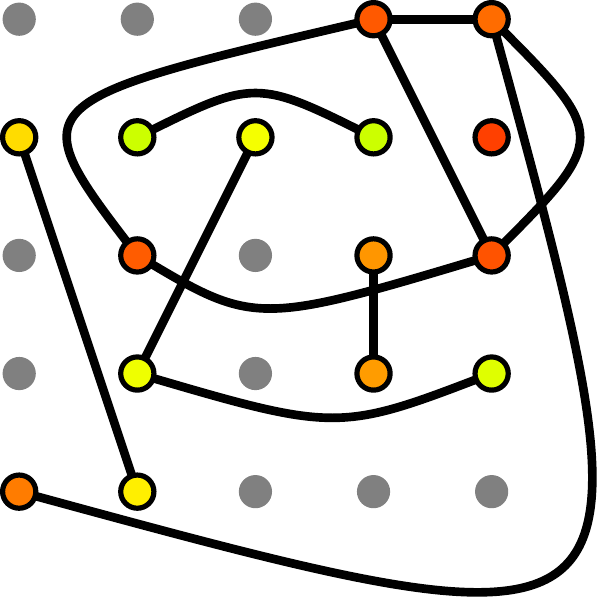}(b)
    \hfill
    \includegraphics[width=0.08\textwidth]{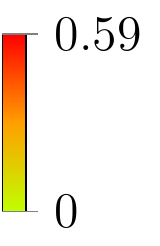}

    \caption{(a) Sketch of a threshold network, defined by Eqn.~\eqref{eq:4} with $\epsilon=0.028$.
      The color code indicates the values of $x_{i}$. At difference with Figure~\ref{fig:2d}, now only agents with similar $x_{i}$ are connected. 
    (b) Sketch of a threshold network with adjusted preferences, defined by Eqn.~\eqref{eq:9}. Note that some links have disappeared, which fragments the network into disconnected components.}
  \label{fig:thresh}
\end{figure}

Evidently, despite the same mean degree the \emph{topologies} of the random network and the threshold network are very different.
For the latter also the values of $x_{i}$ matter, which are ignored in the random model.
Figure~\ref{fig:thresh}(a), which should be compared to Figure~\ref{fig:2d}(b) illustrates this. 

\subsection{Dynamics on the threshold network}
\label{sec:dynam-thresh-netw}

To enrich the hit and flop dynamics, which entirely depends on the percolation density $\rho_{c}$, the SP model introduces also a dynamics of the agent's preferences, i.e. $x_{i}$ becomes time dependent. 
If agent $i$ liked the last movie, i.e. if $x_{i}(t-1)\leq q$, then this agent raises its preference for the next one by a fixed value $\delta x$.
Conversely, if the agent did not like the last movie, $q<x_{i}$, it lowers the value of the preference by $\delta x$.
This models an adaptation process with respect to the movie quality $q$:
\begin{align}
  \label{eq:8}
  x_{i}(t)=&x_{i}(t-1)+\delta x \quad \mathrm{if} \quad x_{i}(t-1)\leq q \nonumber \\ 
  x_{i}(t)=&x_{i}(t-1)-\delta x \quad \mathrm{if} \quad x_{i}(t-1) > q 
\end{align}
As a result, the distribution of the agents preferences, $U(x)$, evolves such that it is narrowly centered around the value of $q$, i.e. $\mean{x}=q$.

A similar dynamic argumentation was used also for the movie quality $q$.
If the last movie was a flop, then the quality of the next movie is increased, it the movie was a hit, the quality is decreased.
We can know about hits and flops only after sufficiently large times $t$, i.e. at the end of a cascade.
To distinguish this, let us introduce a larger time scale $T$, at which the movie quality is adapted:
\begin{align}
  \label{eq:81}
  q(T)=&q(T-1)+\delta q \quad \mathrm{if} \quad X(T-1) \approx 0 \nonumber \\ 
  q(T)=&q(T-1)-\delta q \quad \mathrm{if} \quad X(T-1)\approx 1
\end{align}
Thus, we have to rerun the dynamics on the percolation cluster consecutively for a larger number of time steps $T$.
For the case $\delta x=0$, $\delta q>0$, i.e. only the movie quality adapts, the SP model results in an equilibrium movie quality $q=q_{c}$, i.e. the quality level converges to the percolation threshold.
If both the agents' preferences and the movie quality are allowed to adapt, $\delta x> 0$, $\delta q>0$, one finds that both variables $x$ and $q$ evolve toward their percolation threshold value.
This case is known in the literature as \emph{self-organized criticality} \citep{10.2307/24936753}.

It is a main conceptual drawback with respect to the social science application that agents \emph{only} respond to the movie quality, but not at all to the social interaction.
This way, again, the existence of a social network is ignored as long as the density of the network is above the percolation density, $\rho_{c}$. 
So, in a next step, we assume that a successful interaction between agents results in an adaptation of their individual preferences.

More specifically, the information exchange is directed and time dependent.
Agent $i$ has seen the movie at time $t$.
Only if $x_{i}(t)\leq q$, it will pass on the information about the movie quality to those agents $j$ that have a link to $i$.
We now assume that agents $j$, because of the interaction with $i$, adjust their preferences $x_{j}$ as follows:
\begin{align}
  \label{eq:9}
  x_{j}(t+1)=x_{j}(t)+\mu \left[x_{i}(t)-x_{j}(t)\right]\ \Theta\left[\Delta z_{ji}\right]
\end{align}
This dynamics resembles the bounded confidence model \citep{hegselmann2002opinion,deffuant2001mixing}, where agents converge with their opinion towards the common mean, as the result of their interactions.
The difference here is that only the agent that \emph{receives} the information about $q$ adjusts its preference $x_{j}$ either to lower or to higher values.
The agent that \emph{sends} this information does \emph{not} also adjust its preference $x_{i}$ at the same time.
It may have changed this value at the time $t-1$ which it received the information from its neighbors.
This unilateral adjustment is in line with more recent opinion dynamics models \citep{Garcia2020}. 

The dynamics of Eqn.~\eqref{eq:9} changes the preference values $x_{j}$ of all agents that receive the information.
This again has consequences for the $\Delta x_{ij}(t)$, which also change, and for the existence of a link.
Because $j$ has adjusted its preference towards $i$, it will no longer be able to interact with another neighbor $k$, because of Eqn.~\eqref{eq:4}.
As we do not assume the formation of new links here, the network will become more sparse as a result of the adjustment dynamics, and can possibly break into smaller disconnected components.
The outcome of such a process is shown in Figure~\ref{fig:thresh}(b).

\section{From threshold networks to adaptive networks}
\label{sec:from-thresh-netw}

\subsection{Two time scales}
\label{sec:two-time-scales}

The previous steps have not considered any dynamics for the network topology.
Once this was created with the given network model, the structure was fixed.
With the last step we have already introduced a mechanism to delete existing links as the result of the agent dynamics.
To prevent the network breakdown into disconnected components, however, we then also need to assume a mechanism for link formation.

\begin{figure}[htbp]
  \includegraphics[width=0.35\textwidth]{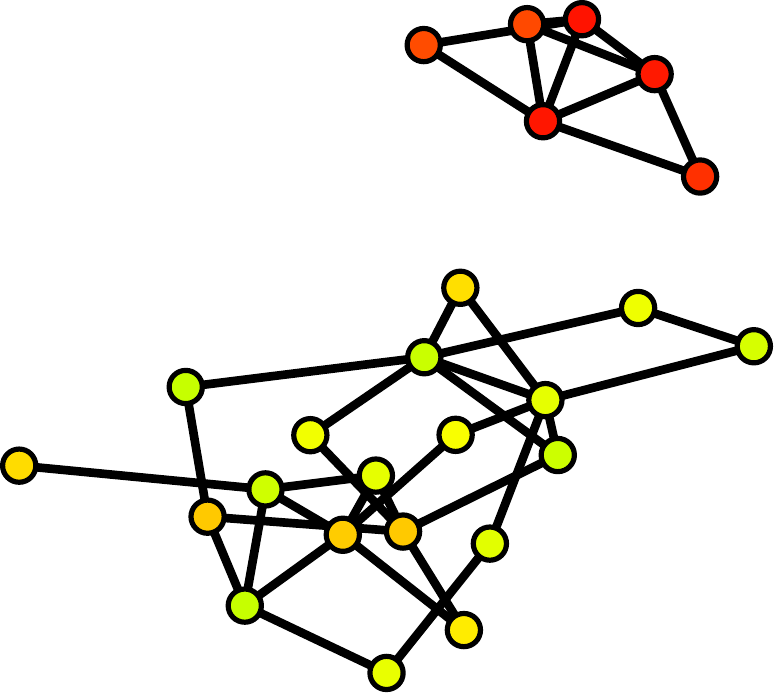}(a)
  \hfill
  \includegraphics[width=0.35\textwidth]{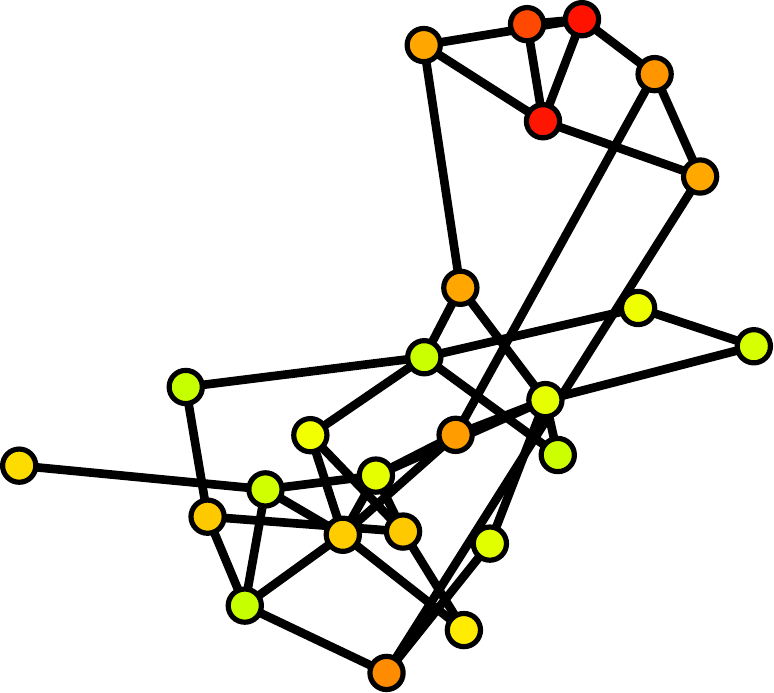}(b)
    \hfill
    \includegraphics[width=0.08\textwidth]{figures/legend}
    \caption{(a) Sketch of an adaptive network with consecutive link formation at time $T$=100.
      (b) Sketch of an adaptive network with consecutive link formation and in-group influence, Eqs.~\eqref{eq:13},\eqref{eq:14} at time $T$=100.
      after   threshold network with adjusted preferences, defined by Eqn.~\eqref{eq:9}. Note the large connected component.
       The color code indicates the values of $x_{i}(T)$ at time $T$. 
    }
  \label{fig:adapt}
\end{figure}

To do so, we make use of the two time scales already mentioned before, in a slightly different manner. 
The shorter time scale $t$ describes the dynamics of the agent preferences, $x_{i}(t)$, Eqn.~\eqref{eq:9}. 
It returns a (quasi)stationary value $x_{i}^{\mathrm{stat}}$ quickly.
The longer time scale $T$ now describes the change of the network topology as a result of the existing values $x_{i}^{\mathrm{stat}}(T)$.
In this modified model, links are not there initially. Instead, at each time step $T$ we select one pair of agents $(i,j)$ and test whether these agents would form a link based on their similar preferences, i.e. based on $z_{ij}(T)$, Eqn.~\eqref{eq:4}.
If $z_{ij}(T)\geq 0$ and no link exists at time $T$, a new link is formed and the information about $q$ is transferred from $i$ to $j$.
If a link already exists and $z_{ij}(T)\geq 0$, it is used for the information transfer.
If the link exists from previous interactions, but  $z_{ij}(T)<0$ at time $T$, the link will be removed.
That means the network topology can constantly adapt to  the current distribution of the $x_{i}(T)$.
Figure~\ref{fig:adapt}(a) shows the outcome of such a process. 

So, from the SP model we have kept the idea to increase or decrease the values of the preferences, $x_{i}(t)$.
But different from the SP model, we do not consider this adjustment in fixed values, but in response to the preference from the senders that have transmitted the information about $q$.
Still, agents have the chance to converge with their preferences to the value needed for the percolation threshold.
However, they do not do this in response to $q$, but because of social interactions.

\subsection{Group dynamics}
\label{sec:group-dynamics}

The network dynamics described above does not necessarily result in a stationary state.
Instead, with every new selection of pairs of agents, links will change.
To introduce an element of stability in this dynamics, we follow a suggestion from \citep{Groeber2009} and consider that agents form \emph{in-groups} that contain those agents they had previously interacted with.
From a socio-economic perspective, interaction is \emph{costly} also because the search for interaction partners is costly.
So, once an agent ``found'' someone with a similar preference, it should try to keep the relation alive, that means, the link alive.

As a consequence, agents do no longer decide about link formation and link deletion based on their actual values $z_{ij}(T)$.
Instead, they take the preferences of their existing in-group $I_{i}(T)$ into account, such that an  \emph{effective preference} is considered: 
\begin{align}
x_i^{\textrm{eff}}(T) =& \left[1-\alpha_i(T)\right]x_i(T) + \alpha_i(T) \mean{x}_i^I(T) 
                   \label{eq:13}
\end{align}
Here $\mean{x}_i^I(T)$ is the mean preference of agents in the in-group of $i$, and $\alpha_{i}(T)$ weights this influence against the ``native'' preference $x_{i}(T)$ of agent $i$, considering the size of the in-group, $\abs{I_{i}(T)}$:
\begin{align}
  \label{eq:14}
  \mean{x}_i^I(T) = & \frac{1}{\abs{I_{i}(T)}}\sum_{j\in I_{i}(T)} x_{j}(T)		\;;\quad \alpha_i(T) = \frac{|I_i(T)|}{|I_i(T)|+1}
\end{align}
While agents adjust their preferences $x_{i}(t)$ still according to Eqn.~\eqref{eq:9}, their effective preferences $x_{i}^{\mathrm{eff}}(T)$ decide about their interactions, i.e.
$z_{ij}^{\mathrm{eff}}(T)= \epsilon - \abs{x^{\mathrm{eff}}_{j}(T)-x^{\mathrm{eff}}_{i}(T)}$.
Only if an interaction takes place, i.e. $z_{ij}^{\mathrm{eff}}(T)\geq 0$, $j$ is added to the in-group of $i$ and a link between agents $i$ and $j$ is formed.

This dynamics is really interesting
because a change of $\mean{x}_{i}^{I}(T)$ can occur even if $i$ does not interact.
This impacts $x_{i}^{\mathrm{eff}}(t)$ continuously.
So, two agents $i$ and $j$ randomly chosen at different times may form a link later, or may remove an existing link because of their  
in-groups' influence, as illustrated in Figure~\ref{fig:adapt}(b).
This feedback between agents' preferences  and their in-group structure sometimes allows to obtain a spanning cluster
even in cases where the original dynamics would fail.

\section{Discussion}
\label{sec:discussion}

The aim of the SP paper \citep{solomon-et-00} was to transfer a physical percolation model into a social context.
This approach to reinterpret known results from physics as social insights was typical for many sociophysics models of that time \citep{schweitzer2018}. 
It does not change the main results of the physical model.
In our case we again find that $q_{c}=\rho_{c}=0.593$, which is the only relevant parameter to quantify the model outcome.

But as I have shown in this short paper, such simplifying physical models still have the potential to bridge the gap toward social phenomena if they are enriched by additional assumptions about social interactions.
These assumptions can often be rooted in social theories, for instance about homophily \citep{flache2017} or cognitive dissonance \citep{Schweitzer2014}.
While this requires some effort, it allows to connect to other disciplines, for example management sciences, where similar problems of product adoption \citep{kiesling2012agent,Parry_2011} and information contagion \cite{Bartal_2020,Min_2018,Centola2007} are addressed. 

This discussion could not be accomplished here.
Instead, I addressed another interesting issue, namely the connection between lattices and networks.
The information transfer between agents is essentially a dynamic process that runs on, and is constraint by, a network with fixed topology.
Thus, if we start from a 2d regular lattice and only consider the percolating cluster, how can this be mapped to a random network?
If the critical density $\rho_{c}$ is known from percolation theory, we can use the $G(n,p)$ model of random networks to determine a link probability $p$ that preserves certain characteristics of the regular-random network formed by the percolating cluster.

One step further, we can also link such networks to threshold networks, in which links are determined by agent quantities, $x_{i}$.
The threshold $\epsilon$ to form a link can, under certain assumptions about the distribution of $x$, be expressed in terms of the link probability $p$.
This nice insight can be utilized to generate networks in different contexts (percolation, random link formation, social affinity) that still share certain network characteristics.

\subsection*{Acknowledgements}
\label{sec:acknowledgement}

I'd like to thank Giacomo Vaccario and Giona Casiraghi for discussions and Jordi Campos for help with the TikZ figures.

\small \setlength{\bibsep}{1pt}

\end{document}